# Emergence of Three-fold Symmetric Helical Photocurrents in Epitaxial Low Twinned Bi$_2$Se$_3$

Blair C. Connelly*, Patrick J. Taylor, George J. de Coster

*U.S. Army DEVCOM Army Research Laboratory, Adelphi, Maryland 20783, USA*

(Dated: March 13$^{th}$, 2023)

*e-mail: blair.c.connelly.civ@army.mil

**Abstract**

We observe enhanced three-fold symmetric helicity-dependent topological photocurrents using time-domain THz spectroscopy in epitaxially-grown Bi$_2$Se$_3$ with reduced crystallographic twinning. It is established how twinned crystal domains introduce competing responses that obscure inherent nonlinear optical responses of the intrinsic crystal structure. Minimizing this defect reveals strong nonlinear optical response currents whose magnitude and direction depend on the alignment of the excitation to the crystal axes and follow the three-fold rotational symmetry of the crystal structure. Notably, the azimuthal dependence of the photoresponse persists for helical excitations – an unprecedented result we attribute to the photon drag effect, where the photon momentum acts as an applied in-plane field that is stationary in the laboratory frame. Additionally, the sign of the resultant THz signal inverts when the helicity of incident light is switched from right to left circularly polarized, indicating a reversal of the photocurrent. Our results demonstrate that even extended domain defects can obscure intrinsic physical processes, making the study of single domain thin films crucial to the observation of phenomena that couple topological order and crystal symmetries.

**Main Text**

Light-matter interactions within topological insulators (TIs) and Weyl semimetals (WSMs) yield unique photoresponses due to spin-momentum locked surface and bulk states, respectively.[1-11] Specifically, nonlinear optical processes such as the photogalvanic effect (PGE) and photon drag effect (PDE) generate strongly polarization-sensitive photocurrents. Several experiments have observed polarization dependent photocurrents in these materials, with WSMs boasting inherently stronger responsivities due to the participation of bulk states in generating photocurrents.[12-14] A large body of research exists on the nature of photocurrents in the $(Bi,Sb)_2(Se,Te)_3$ (BSST) family of TIs.[15-17] These studies span all-optical contactless probing of photocurrents using ultrafast time-domain THz spectroscopy (TDTS) and direct measurement of photocurrents in fabricated devices.[18-25] For BSST TIs, the surface photocurrents generated by PGE and PDE are expected to reflect the underlying three-fold rotational symmetries of the crystal, and propagate in opposite directions when generated by left versus right circularly polarized light.[8,9] These observations are strikingly absent in epitaxially-grown thin film samples, but have been seen in samples derived from single crystals,[1] where a current reversal and large responsivities were engineered through Fermi level control[16] or utilizing the photothermal effect.[5] In this article, we posit that the traditionally weaker response of thin films grown via molecular beam epitaxy (MBE) is due to crystallographic twin defects and present theory showing how their presence obscures the underlying crystal symmetry and related polarization-dependent photoresponses.[17,26]

We present TDTS measurement and analysis of high-quality $Bi_2Se_3$ – grown to preferentially obtain a single twinning domain to significantly reduce twinning defects – which demonstrate a colossal enhancement in helicity dependent photocurrents over highly twinned samples. Additionally, we report on the emergence of three-fold symmetry in polarization dependent photocurrents, as well as a clear demonstration of the dependence of the photocurrent direction on helicity and plane-of-incidence. Reduction of this obfuscating effect thus allows for the observation of fundamental physical responses that have not been previously reported, revealing the twinning defect to be particularly deleterious to the expression of phenomena inherent to the topological nature of the underlying material. These nonlinear and topological photoresponses subsequently demonstrate strongly azimuthally-dependent signatures that can be leveraged in device designs such that the full polarization state of incident light can be measured. Intricate knowledge of these azimuthally and polarization-dependent photoresponses can inform clever contact schemes that enable the complete decoding of the degree of ellipticity *and* helicity for polarimetric measurement of light.

$Bi_2Se_3$ grows in quintuple layers of $Se^I$-$Bi$-$Se^{II}$-$Bi$-$Se^I$ that stack in the growth direction $\hat{z}$, with van der Waals bonding between neighboring $Se^I$ atoms of the quintuple layers. A perfect $Bi_2Se_3$ crystal is symmetric under the operations of the point group $D_{3d}$, which is generated by three-fold rotations about the z-axis, $C_3$, inversion, $i$, and mirror reflection along the (100) x-axis, $\sigma_{\hat{x}}$. The $Bi_2Se_3$ lattice can be 'twinned' by mirroring it along the (010) y-axis. A top-down view of the final Se-Bi layers of a particular choice of 'twin domain' is presented in Fig. 1a. MBE growth often allows different twinning domains of $Bi_2Se_3$ to percolate randomly on a substrate, resulting in a material that effectively has an additional mirror reflection symmetry along the y-axis, inducing a six-fold symmetry in addition to the three-fold symmetry of perfect $Bi_2Se_3$. In this case, the point group describing the effective crystal is $D_{6h}$ and certain three-fold symmetric responses will be suppressed in measurements of these thin-film samples (see Supplemental Material).[19] Judicious substrate choice and growth conditions enable a single domain of $Bi_2Se_3$ to dominate in MBE growth.[26-30] For this study, a 12.4-nm-thick epitaxial film of $Bi_2Se_3$ was grown via molecular beam epitaxy on a semi-insulating InP(111)B substrate with a vicinal mis-orientation of two degrees (2º) toward the <010> direction at a substrate temperature of 290 °C. Reflection high-energy electron diffraction (RHEED) and X-ray diffraction spectra confirm high-quality epitaxy with an approximate twin domain ratio of 5:1 (see Figs. S1-S3 in the Supplemental Material). An in-situ capping layer of ~10 nm of cubic EuS was deposited to prevent environmental degradation of the surface states in atmosphere.[25] The large bulk energy gap of 3.1 eV makes the EuS layer transparent to the 1.55-eV excitation laser. Additionally, at room temperature EuS on $Bi_2Se_3$ has been conclusively demonstrated to be a nonmagnetic dielectric and therefore no intrinsic time-reversal symmetry breaking effects are anticipated to contribute to the experiment.[31]

In this article, we present polarization-sensitive TDTS measurements at room temperature in an open-air laboratory environment of photocurrent responses in a low-twinned $Bi_2Se_3$ thin film. Corresponding data for a high-twinned sample is presented in the Supplemental Material for comparison. The measurement technique is an ideal way to extract photocurrent physics directly and non-destructively without interferences from photothermal currents, whilst maintaining sensitivity to directionality, helicity and amplitude. As shown in Fig. 1a, the ellipticity of an 800-nm excitation laser pulse (~125-fs pulse width, ~5-μJ pulse energy), is varied between linear horizontal, in the *yz*-plane, and right- and left-circular polarizations (RCP and LCP, respectively) using a quarter wave plate, QWP, and incident upon the sample at a 45° angle of incidence. Here, polarizations of RCP, linear and LCP are obtained with a QWP angle, $\theta$, of 45°, 90° and 135°, respectively. Excitations within the bulk and between the 1st and 2nd topological surface states[1] generate photocurrents in the sample, where the time-rate-change of the current, $\frac{d}{dt}j_{i=x,y,z}$, is proportional to the emitted THz electric field, $S_{i=x,y,z}$. The time-dependence of the emitted THz is detected using electrooptic sampling on a ZnTe crystal,[32] and the signal's horizontal and vertical components are distinguished using a wire grid polarizer (WGP) with 99.9% rejection of the cross-polarized THz signal. The horizontal component, $S_{yz}$, combines surface-parallel and perpendicular currents in the sample's $yz$-plane. The vertical component, $S_x$, is purely due to a surface-parallel current in the $x$-direction, which is orthogonal to the laser excitation's plane of incidence.

To interrogate crystal symmetries, we explore the dependence of $S_{yz}$ on the sample's azimuthal angle, $\phi$, for a given input polarization. Here, $\phi = 0°$ is nominally represented by the top view of the sample in the $xy$-plane in the inset of Fig. 1a; increasing $\phi$ indicates counterclockwise rotations of the sample, as shown. Figure 1b plots a series of time-dependent THz waveforms from $\phi = 0°$ to $\phi = 60°$, under linear horizontal polarization excitation ($\theta = 90°$), and Fig. 1c plots the peak-to-peak signal amplitude over a full 360° sample rotation. A clear three-fold symmetry emerges and the data is well fit by a $\cos(3\phi + \delta)$ dependence where $\delta = 6°$ defines the (100) axis.

To interpret the physical processes observed in the THz data, we use symmetry analysis[33] to determine the 2$^{nd}$ order nonlinear optical (NLO) contributions to the photocurrent density, $j_P$, which can be decomposed into circular (C) and linear (L) PGE and PDE terms, $\boldsymbol{j}_P = \boldsymbol{j}_{CPGE} + \boldsymbol{j}_{LPGE} + \boldsymbol{j}_{CPDE} + \boldsymbol{j}_{LPDE}$. Explicitly presented in the Supplemental Material, we find the dependence of the PGE and PDE contributions to $\boldsymbol{j}_P$ as a function of $\theta$ and $\phi$ in our experiment:

$$\boldsymbol{j}_{LPGE} = \begin{pmatrix} \eta_{x4} \sin 4\theta - \sin 3\phi \left(\tilde{\delta} + \tilde{\kappa}_4 \cos 4\theta\right) + \tilde{\eta}_4 \cos 3\phi \sin 4\theta \\ \delta_y + \kappa_{y4} \cos 4\theta + \cos 3\phi \left(\tilde{\delta} + \tilde{\kappa}_4 \cos 4\theta\right) - \tilde{\eta}_4 \sin 3\phi \sin 4\theta \\ \delta_z + \kappa_{z4} \cos 4\theta \end{pmatrix}, \quad (1a)$$

$$\frac{\boldsymbol{j}_{LPDE}}{q} = \begin{pmatrix} \xi_{x4} \sin 4\theta + \sin 3\phi \left(\tilde{\Delta} + \tilde{\lambda}_4 \cos 4\theta\right) + \tilde{\xi}_4 \cos 3\phi \sin 4\theta \\ \Delta_y + \lambda_{y4} \cos 4\theta + \cos 3\phi \left(\tilde{\Delta} + \tilde{\lambda}_4 \cos 4\theta\right) - \tilde{\xi}_4 \sin 3\phi \sin 4\theta \\ \Delta_z + \lambda_{z4} \cos 4\theta + \cos 3\phi \left(\tilde{\Delta}_z + \tilde{\lambda}_{z4} \cos 4\theta\right) + \tilde{\xi}_{z4} \sin 3\phi \sin 4\theta \end{pmatrix}, \quad (1b)$$

$$\boldsymbol{j}_{CPGE} = i \sin 2\theta \, (\eta_2, 0, 0)^T, \quad (2a)$$
$$\boldsymbol{j}_{CPDE} = i q \sin 2\theta \left(\xi_2 + \tilde{\xi}_2 \cos 3\phi, -\tilde{\xi}_2 \sin 3\phi, 0\right)^T. \quad (2b)$$

Here $q$ is the incoming light's momentum, and the $\eta, \kappa, \xi, \lambda, \delta$ and $\Delta$ coefficients are related to elements of the relevant NLO response tensors and incident electric field, and are defined in the Supplemental Material. The circular PGE and circular PDE (CPGE and CPDE) and linear PGE and linear PDE (LPGE and LPDE) are distinguished by their $2\theta$ and $4\theta$-dependence on the QWP angle, respectively. Dependence of $\boldsymbol{j_P}$ on frequency and momentum is implicit in the coefficients. We note that our analysis differs from previous works by not taking the DC limit of $\boldsymbol{j_P}$ as our photocurrents are strongly transient, which in turn imposes fewer constraints on the elements of the NLO response tensors.[1] The analysis reveals that PGE contributions to $\boldsymbol{j_P}$ can only arise from the inversion breaking surface of Bi$_2$Se$_3$, as the third rank tensor governing the PGE response is zero for the inversion symmetric bulk. Conversely, the fourth rank tensor for PDE is invariant under inversion, so both bulk and surface states can contribute to PDE.

When considering linearly polarized light, *i.e.*, $\theta = 0°$ or $90°$, Eq. (1) simplifies to show a $\cos 3\phi$ periodicity in $j_y$ and $j_z$. This periodicity is transferred to $S_{yz}$ as Braun *et al.*[1] show $S_{yz} \propto S_y + \alpha\, S_z$, where $\alpha \approx 0.3$ a weighting term accounting for geometry and index of refraction at THz frequencies. The $\cos 3\phi$ dependence of $S_{yz}$ in Figs. 1b and 1c is thus captured by the NLO symmetry analysis. In highly twinned samples, signatures of three-fold periodicity are diminished as competing twin domains and contribute $\tilde{\xi}_2$ terms with opposite sign, leading to a near net cancellation (see Fig. S4).[19]

Figure 1d shows the $S_x$ component of the THz waveforms for RCP (blue), linear-horizontal (black) and LCP (red) polarized light for sample azimuthal angles $0°$ (solid) and $180°$ (dashed). As the CPGE response can only originate from the inversion breaking surface, and it is predominantly due to the topological surface state,[21] the related photocurrent propagates orthogonal to the plane of incidence and is manifest in the $S_x$ component of the emitted THz. From Fig. 1d we can see that in low-twinned Bi2Se3 there is a nearly complete reversal of the helical photocurrent when the incident light is switched from RCP to LCP light, creating counter-propagating helical photocurrents. Additionally, we see that as the sample is rotated 180°, the helical photocurrents reverse direction. This reversal with crystal rotation was seen to a muted effect in earlier experiments[19,34] and indicates that intrinsic crystal symmetries impact the helical photocurrents. Equation (2) shows $j_{CPGE,x}$ is $\phi$-independent, while $j_{CPDE,x} \propto \xi_2 + \tilde{\xi}_2 \cos 3\phi$, and can change signs every $\Delta\phi = 60°$ provided $\tilde{\xi}_2 > \xi_2$ and minimal $j_{LPDE,x}$. Additionally, for $\theta \in \{45°, 135°\}$ and $\phi = 0°$ and $180°$ Eq. (1) shows $j_{LPGE,x} = j_{LPDE,x} = 0$. Therefore, CPDE and careful choice of plane of incidence can drive the current reversal in Fig. 1d.

The presence of three-fold symmetry in $S_{yz}$ and current reversal in $S_x$ motivates the investigation of the impact of crystalline symmetries on the $S_x$ data. In Fig. 2 we present a series of $S_x$ data taken at every 15° of azimuthal sample orientation for RCP, horizontal linear polarization and LCP. Note, at these three QWP angles ($\theta = 45°, 90°, 135°$) we have $\sin 4\theta = 0$ and so only $\sin 2\theta$ and $\cos 4\theta$ dependent terms in Eqs. (1) and (2) contribute. The three-fold symmetry of the $S_x$ data is easily observed in Fig. 2, wherein waveforms every 120° of sample rotation are identically colored for emphasis. This is represented in Fig. 3a as a radial plot of the $S_x$ waveforms for $\theta = 90°$, with time on the radial axis and $\phi$ on the polar axis; analogous radial plots for the RCP and LCP data are provided in the Supplemental Material. Figures 2a and c show the sign of $S_x$ for RCP and LCP light reverses every $\Delta\phi = 60°$, which is consistent with the discussion in the previous paragraph and indicates a large $\tilde{\xi}_2$ coefficient in $j_{CPDE,x}$. We observe that the largest helical photocurrents occur for sample orientations $\phi = 60°, 180°$ and $300°$, which corresponds to angles where the linear photocurrents are also minimized. At these angles the incoming light is anti-parallel to the ultrafast shift current that transfers charge along the Se—Bi bond[1,36] (as shown in Fig. 1a) and correspondingly $S_{yz}$ is minimized in Fig. 1b. This demonstrates that particular sample orientations allow for the observation of helical photocurrent responses. Conversely, a 60° azimuthal rotation leads to a dominant shift current contribution, as the detection is aligned with the Se—Bi bond.

We analyze the azimuthal rotational symmetries of $S_x$ by computing the waveforms' transforms with respect to $\phi$. One can express $S_x$ as a Fourier series $S_x(t, \phi, \theta) = \sum_{n \in \mathbb{N}_0} c_n(t, \theta) \cos(n\phi) + s_n(t, \theta) \sin(n\phi)$ with the Fourier coefficients formally determined by $c_n(t, \theta) + i\, s_n(t, \theta) \equiv \frac{1}{\pi} \int_0^{2\pi} S_x(t, \phi, \theta)\, e^{in\phi}\, d\phi$ and $c_0 = \frac{1}{2\pi} \int_0^{2\pi} S_x(t, \phi, \theta)\, d\phi$, where we have accounted for the offset in the azimuthal angle identified by the $S_{yz}$ data fit in Fig. 1b. Given the finite nature of the data set, we use third order piecewise polynomials to interpolate the data taken for $\phi \in [0°, 360°)$, and then use numerical integration to obtain $c_n \equiv c_n(t, \theta)$ and $s_n \equiv s_n(t, \theta)$, which quantify the degree of $\cos n\phi$ and $\sin n\phi$ periodicity present in a waveform at a time $t$ and polarization $\theta$. The three-fold $(c_3, s_3)$, one-fold $(c_1, s_1)$ and azimuthally independent $(c_0)$ coefficients were found to be the dominant contributions to $S_x$ and are plotted in Figs. 3b and 3c. The transformed data shows that for RCP/LCP light the three-fold coefficients obey $c_3(t, 45°) \cdot c_3(t, 135°) \leq 0$ and $s_3(t, 45°) \cdot s_3(t, 135°) \geq 0$. This behavior is perfectly captured by the NLO response current in Eq. (2b), which simplifies for RCP/LCP light to $j_x \equiv c_0^{\pm} + s_3 \sin 3\phi \pm c_3 \cos 3\phi$, for $c_0^{\pm}, s_3$ and $c_3$ time dependent coefficients. Figure 3c shows that the coefficients $c_0$ have opposite signs for LCP and RCP light and are smaller than $c_3$ and/or $s_3$. This suggests the CPGE contribution to the photocurrent in Eq. (2a) is smaller than the one from CPDE in Eq. (2b) and is why we see such a pronounced three-fold symmetry in the data.

Figure 3c further shows there is a non-trivial one-fold periodic contribution from $s_1$ and $c_1$ to $S_x$, which is not captured by the symmetry analysis where perfect planar crystals is assumed. The one-fold aberration may be induced by the offcut substrate, or as other works found, strain effects in thin films of $Bi_2Se_3$[15] and wandering of the excitation/collection area with sample rotation.[37] It is surprising that the one-fold contribution outweighs the azimuthally independent one, suggesting that if $c_3$ and $s_3$ were weak the directional dependence of the photocurrent on RCP and LCP light could be obscured. Twinned $Bi_2Se_3$ has an effective six-fold rotational symmetry in which case symmetry analysis predicts $s_3 \approx c_3 \approx 0$ (see Supplemental Material). The coupled effects of twinning, small CPGE and one-fold aberration provide a plausible explanation for nonuniversal observation of current reversal in MBE-grown $Bi_2Se_3$.

To explore the full polarization dependence of the photocurrents, $S_x$ waveforms are captured at discrete sample orientations, and at 15° increments of the QWP angle $\theta$ from 0° to 180°; $S_x$ waveforms are duplicated for mathematically equivalent QWP angles from 180° to 360° (see Supplemental Material Eq. (S3)). Contributions from the substrate to the photocurrent have been ruled out through comparison of complementary $S_x$ data, where emission from a bare InP(111) substrate are observed to be an order-of-magnitude smaller (additional details are presented in the Supplemental Material).

Figure 4 presents the low-twinned $Bi_2Se_3$ $S_x$ data with time on the radial axis and $\theta$ on the polar axis. Sample angles $\phi = 0°, 90°, 180°, 240°, 270°$ and 300° were chosen to visualize the three-fold azimuthal symmetry in the mirrored pairs (0°, 240°) and (180°, 300°), and the suppression of CPGE and CPDE in the orientations 90° and 270°. At $\phi = 90°$ and 270° where the strongest linear (i.e., $\theta = 90°$) response was observed in Fig. 2, we see a strong four-fold symmetry in the data, which presents itself as four positive/negative lobes over a 360° rotation in $\theta$. Moreover, the current is seen to reverse direction with a $\Delta\phi = 180°$ sample rotation. At $\phi = 0°, 180°, 240°$ and $300°$ a strong two-fold modulation of the signal in $\theta$ and a sign/photocurrent reversal when switching between RCP and LCP ($\theta = 45°$ and $135°$) is observed, indicating a dominant helical photocurrent response. Once again, the photocurrent is seen to reverse upon a $\Delta\phi = 180°$ sample rotation, which is most evident at $\theta = 60°$ where the observed photocurrents are largest. Additionally, when the $S_x$ waveform is predominantly a function of $\sin 2\theta$, a photocurrent reversal manifests itself as a $\theta = 90°$ rotation between $\Delta\phi = 180°$ pairs ($i.e. -\sin 2\theta = \sin 2(\theta + \frac{\pi}{2})$). Consistent with our analysis of Fig. 2, the strongest two-fold modulation with respect to $\theta$ is observed at $\phi = 180°$ and $300°$.

To analyze the polarization dependent response of the photocurrent we compute the discrete Fourier transforms with respect to $\theta$, where $S_x(t, \phi, \theta) = \sum_{\tilde{\theta} \in \mathbb{N}_0} \tilde{c}_{\tilde{\theta}}(t, \phi) \cos(\tilde{\theta}\theta) + \tilde{s}_{\tilde{\theta}}(t, \phi) \sin(\tilde{\theta}\theta)$. Figure 5 presents the physically significant coefficients $\tilde{c}_4(t, \phi)$, $\tilde{s}_4(t, \phi)$ and $\tilde{s}_2(t, \phi)$ computed by numerically integrating the interpolated $\theta$-dependent data in the same way as for the $\phi$-dependent data in Fig 3. Note, the $\theta \in [0°, 180°]$ measurement range implies we can only calculate even periodic modes over $\theta \in [0°, 360°]$. In this manner we have deconvolved the different polarization dependent responses to isolate the helical and linear photocurrent channels for a given sample orientation. Several observations of the data in Fig. 5 are elegantly matched by the earlier theoretical analysis. For example, $\tilde{s}_4(t, \phi)$ (Fig. 5a) attains its maximum absolute value for $\phi = 180°$ and $300°$, which follows from the functional form $\tilde{s}_4(t, \phi) \sim const. - \cos 3\phi$ imposed by Eq. (1). Additionally, Eq. (1) dictates $\tilde{c}_4(t, \phi) \sim \sin 3\phi$, which is represented in Fig. 5b wherein $\tilde{c}_4(t, \phi = 90°)$ and $\tilde{c}_4(t, \phi = 270°)$ are the largest signal components and have opposite signs.

Figure 5c reveals one of the most impactful results of this study: the helical photocurrent $\tilde{s}_2(t,\phi)$ is dominated by the CPDE contribution. Using Eq. (2) one can show that $\tilde{s}_2(t,\phi) = i(\eta_2 + q\xi_2 + q\hat{\xi}_2 \cos 3\phi) \sin 2\theta$, where $\eta_2$ is the CPGE coefficient and $\xi_2$ and $\hat{\xi}_2$ are the CPDE coefficients. If CPGE were the dominant photocurrent mechanism, $\tilde{s}_2(t,\phi)$ would be azimuthally independent. Instead, we see $\tilde{s}_2(t,\phi)$ attains maximum values at $\phi = 60°, 180°$ and $300°$, mostly obeying three-fold rotational symmetry. The small one-fold $\phi$-dependence discussed following Fig. 4c accounts for a directional enhancement along $180°$: $|\tilde{s}_2(t,\phi = 180°)| \gtrsim |\tilde{s}_2(t,\phi = 300°)|$. If one considers the PDE and PGE response of fully twinned Bi$_2$Se$_3$, which acquires an additional mirror symmetry, all azimuthal dependence drops out of the photocurrents. Therefore, it is *essential* to minimize twinning (as is the case for judicious growth or exfoliation) to see this result.

In this article, we have presented *the first* observation of clear three-fold azimuthal symmetry of photocurrents in TIs that reverse direction for different helicities of light on low twinned MBE grown samples. Crucially, these all-optical measurements ensure that photothermal currents did not impact helical photocurrent reversal.[5,12,37] While weak three-fold azimuthal dependence has previously been reported in the literature, it was only seen as a small modulation on top of a substrate-generated one-fold periodicity, and was not accompanied by helical current reversal.[1,19] We establish a key explanation for the lack of this observation: the presence of twinning defects in MBE-grown TIs, which on average endow the crystal structure with an effective six-fold rotational symmetry. The NLO tensor symmetry analysis in this case is presented in the Supplemental Material and we find that all dependence of the surface current on the azimuthal angle drops out, *i.e.*, $j_x \sim i(\eta_2 + \xi_2) \sin 2\theta + (\eta_{x4} + \xi_{x4}) \sin 4\theta$. Recalling the discussion of Fig. 3, we found that the intrinsic azimuthally independent contributions to the photocurrent were weaker than extrinsic one-fold contributions, which could obscure helical photocurrent reversal in highly twinned materials.

We emphasize that the three-fold periodicity originates from CPDE as the $j_{CPGE,x}$ photocurrent is inherently azimuthally independent.[21] We suspect multiple effects conspire to provide large CPDE-driven responses in our sample(s). First, given that CPDE comes from a fourth rank tensor, it contains bulk crystal contributions, as even rank tensors do not vanish under inversion symmetry, whereas CPGE can only come from an inversion breaking interface. Secondly, it has been well reported that the helical photocurrents that arise in $Bi_2Se_3$ pumped by 800-nm light come from excitations from initial states in the principal Dirac TSS to a 1.7-eV higher energy Dirac TSS[34,35] with opposite chirality[38] in the conduction bands.[39,40] Since photoexcitation with RCP or LCP light requires the final and initial states to have opposite quantum numbers, the spin-flip transition rule can be frustrated in a direct transition if the excitation energy is smaller than the Dirac point separation. This is the case for our 1.55-eV excitation pulse, and so an additional momentum transfer to the electron may be necessary to satisfy the selection rules, i.e. PDE.[41-44] We anticipate that by changing the excitation wavelength and Fermi level in low twinned BSST TIs – thereby tuning between inter- and intra-cone (and sub bulk bandgap) excitations – one can modulate the strength, directionality and clarity of the three-fold symmetries, achieving complete control over TI photocurrents.

**Acknowledgements**

We would like to thank D. Hsieh, D. Rees, M. Khajavikhan and A. Llopis for helpful conversations. This work was funded by an ARAP grant from the OUSD R&E.

**Author contributions**

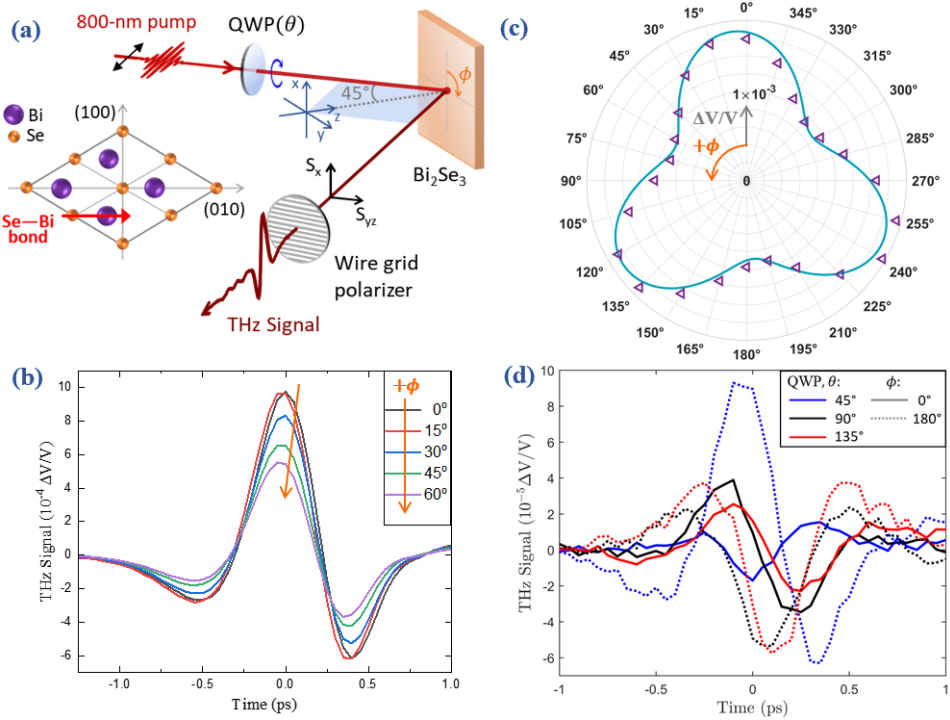

**Figure 1 | Experimental setup and bulk crystal symmetries (a)** Time-domain experiment: An 800-nm, ~125-fs laser pulse is prepared in linear horizontal polarization and incident on a $Bi_2Se_3$ sample at an angle of incidence of 45°; the ellipticity of the light is controlled using a quarter-wave plate (QWP). The emitted THz is collected with 90° off-axis parabolic mirrors and focused on a ZnTe crystal for detection using electro-optic sampling (not shown here). A wire grid polarizer is used to filter the emitted polarization for detection in the horizontal and vertical directions ($S_{yz}$ and $S_x$, respectively). (**a**, *Inset*) Top view of outer most Se and Bi layers of the $Bi_2Se_3$ crystal structure, oriented with an azimuthal angle, $\phi$, of 0°, where the (100) axis is oriented vertically and the (010) axis horizontally. Note: the direction of the shift current occurs along the Se—Bi bond (red arrow). **(b)** Differential THz signals emitted with $S_{yz}$ polarization for displayed $\phi$ under excitation with linear horizontal polarization; the direction of increasing $\phi$ is indicated. **(c)** Peak-to-peak amplitude of $S_{yz}$ as a function of $\phi$ (purple triangles); the fitted function $\propto \cos(3\phi)$ (blue line) demonstrates the three-fold crystallographic symmetry. **(d)** THz waveforms of $S_x$ at $\phi = 0°$ (solid lines) and $\phi = 180°$ (dashed lines) for QWP angles $\theta = 45°$ (right-hand circular; blue), $\theta = 90°$ (linear horizontal; black) and $\theta = 135°$ (left-hand circular; red). Note, both an inversion of sign between excitations of right and left-circular polarization and for sample rotations of 180°.

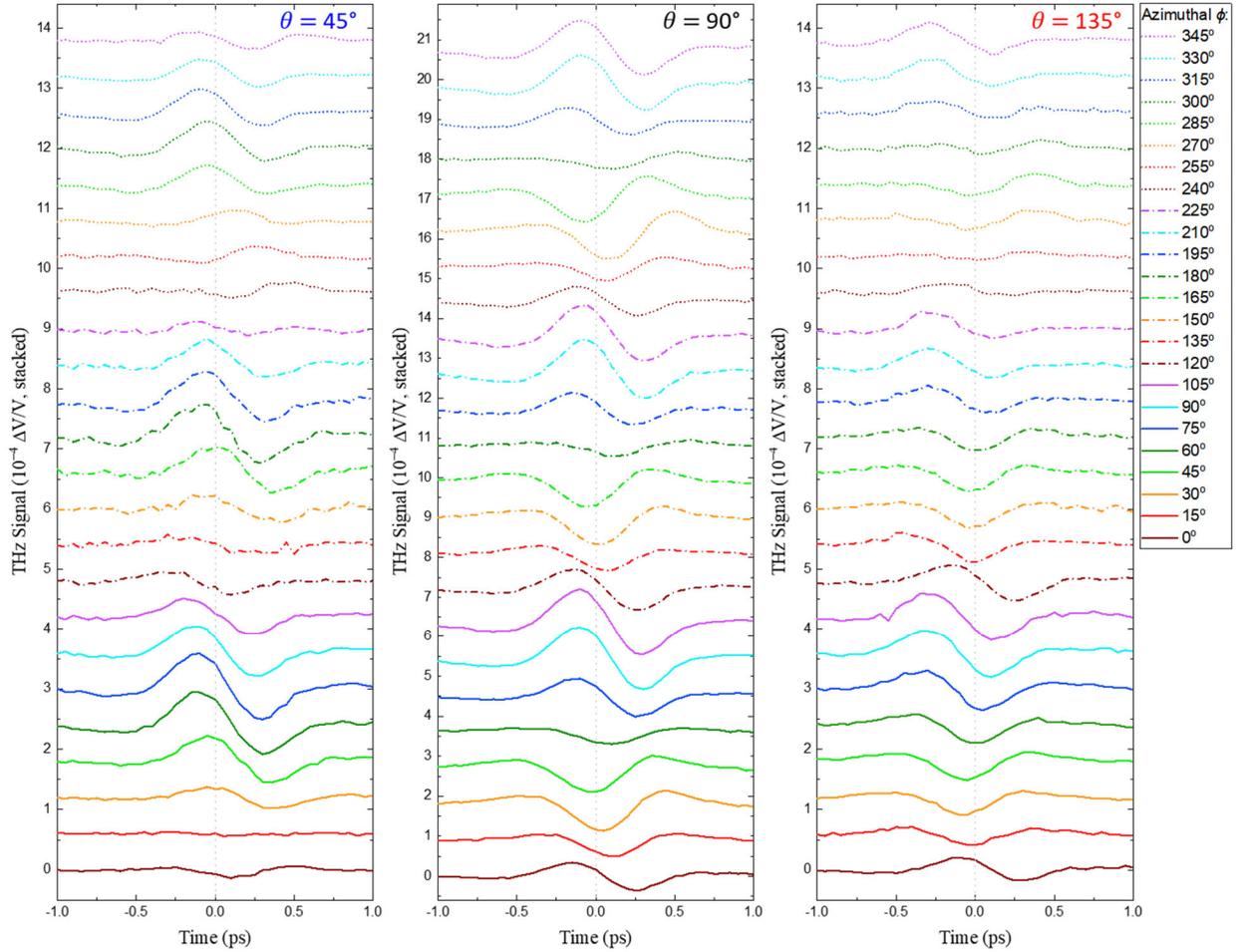

**Figure 2 | Azimuthal dependence of surface parallel photocurrents.** Vertical/surface THz signal, $S_x$, as a function of azimuthal angle, $\phi$, under excitation with: **(a)** Right-handed circular polarization (RHCP, $\theta = 45°$); **(b)** Linear horizontal polarization (Horizontal, $\theta = 90°$); **(c)** Left-handed circular polarization (LHCP, $\theta = 135°$). Three-fold symmetric angles with $\Delta\phi = 120°$ are indicated by the same color. Radial plots can be found in the supplemental information. The reversal of right/left surface currents as a function of azimuthal angle is most evident when comparing 0° to 60°. Notably, at 60° the linear current is minimized, suggesting this crystal axis would be beneficial for binary circularly polarized light detection.

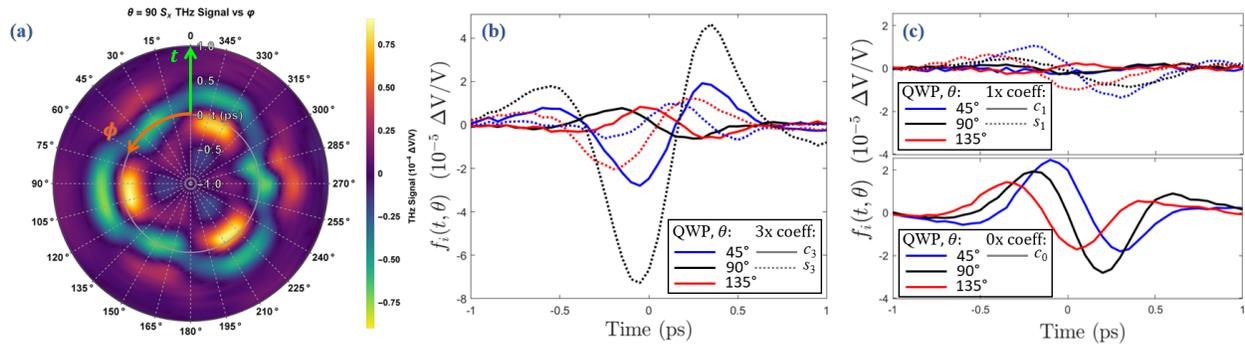

**Figure 3 | Azimuthal angle dependent Fourier decomposed THz contributions. (a)** Data from Fig. 2b as a radial plot of the $S_x$ waveforms for $\theta = 90°$, with time on the radial axis and $\phi$ on the polar axis to visualize the three-fold symmetric response. For each incident polarization in Fig. 2, the THz signal's $\phi$ dependence is separated into time-dependent Fourier series sine and cosine coefficients ($s_n$ and $c_n$, respectively), where **(b)** plots the resultant waveform of the three-fold contributions and **(c)** plots the one-fold (top) and constant (bottom) contributions.

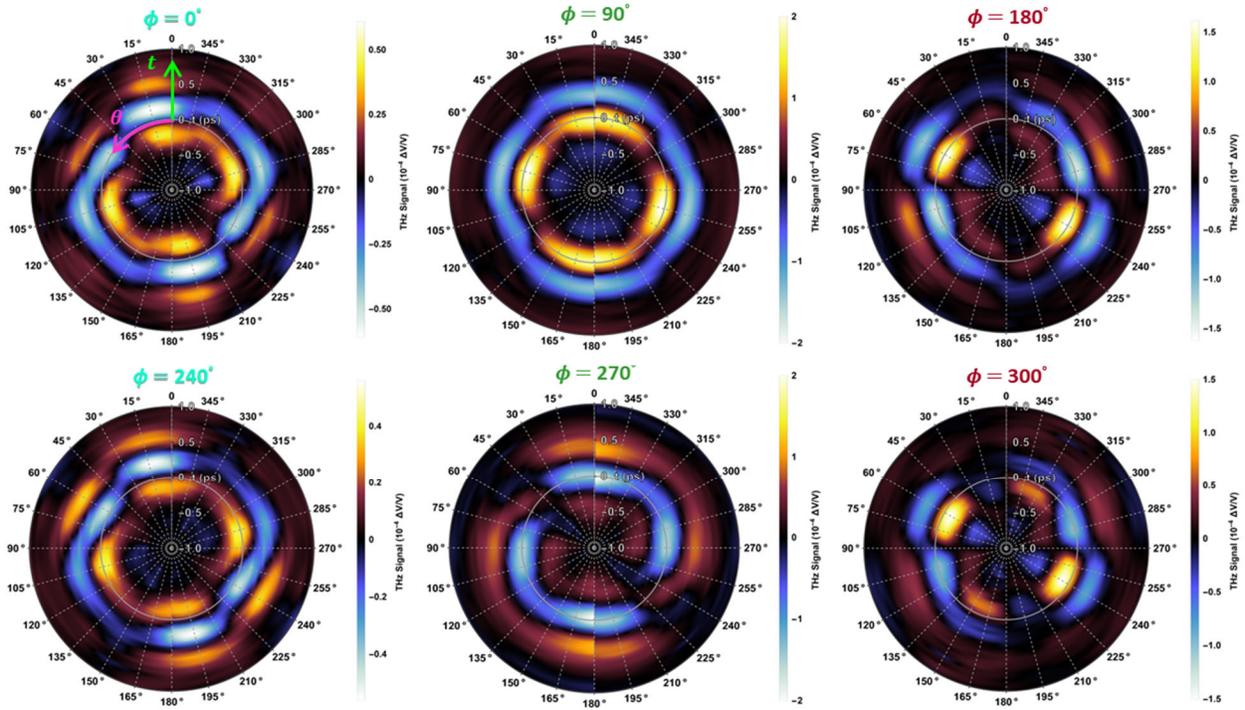

**Figure 4 | Helicity dependence of surface parallel photocurrents.** Radial plots of $S_x$ waveforms as a function of the QWP angle, $\theta$, with time on the radial axis and $\theta$ on the polar axis; each radial plot is displayed on an independent color scale (as shown). The raw data is taken for QWP angles $\theta = 0°$ through $180°$ in $15°$ steps, and then piecewise interpolated with a third-degree polynomial to achieve a finer plot resolution. The interpolated data set is then repeated for $\theta \in (180°, 360°]$ to achieve a full radial plot (QWP polarization states are inherently two-fold periodic). Pairs of three-fold symmetric azimuthal angles (separated by $120°$) are shown in the left column for $\phi = 0°$ and $240°$ and right column for $\phi = 180°$ and $300°$ to demonstrate similarities between signals and two-fold dependence on input polarization. The center column plots $\phi = 90°$ and $270°$, where detection is aligned and counter-aligned with the Se—Bi bond direction, respectively, and strong four-fold, linear dependence on the input polarization is observed.

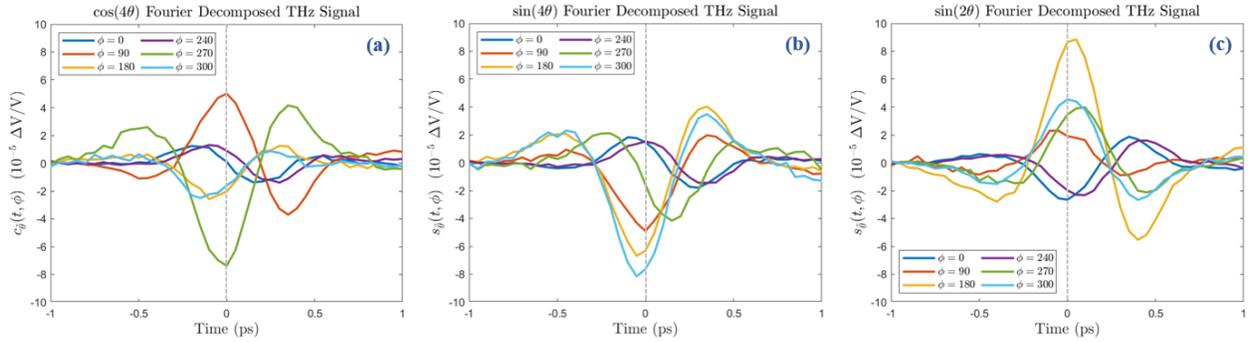

**Figure 5 | QWP angle dependent Fourier decomposed THz contributions.** The time dependent Fourier coefficients corresponding to (**a**) $\cos(4\theta)$, (**b**) $\sin(4\theta)$ and (**c**) $\sin(2\theta)$ are displayed for the same azimuthal angles as Fig. 4. Observe that pairs of coefficients separated by $\Delta\phi = 120°$ show remarkable similarity, highlighting the three-fold rotational symmetry. Deviations from three-fold can be accounted for by the presence of substrate induced one-fold periodic corrections to the photocurrent-azimuthal angle relationships. The strength of the helical photocurrent is governed by the $\sin(2\theta)$ coefficient, which reaches its maximum value at $\phi = 180°$.

**Supplemental Material**

1. Characterization of low twinned $Bi_2Se_3$ thin films

2. Comparison of THz data on low and high twinned $Bi_2Se_3$ samples

3. Symmetry analysis of NLO tensors

4. Azimuthal dependence of surface parallel photocurrents

5. Extended Fourier Analysis

6. THz Waveforms for each QWP angle

7. InP Data

**1. Characterization of low twinned $Bi_2Se_3$ thin films**

The evidence for the reduction in the concentration of twin defects can be found in structural analysis of the obtained epitaxial heterostructure. That analysis includes reflection high-energy electron diffraction (RHEED) in Fig. S1 and high-resolution x-ray diffraction (HR-XRD) pole-scans in Figs. S2 and S3. The reduction of twin density is indicated by the difference in intensity of diffraction spots observed during the deposition of a well-behaved cubic layer on top of the $Bi_2Se_3$. The cubic layer (europium sulfide, EuS) begins epitaxy on the $Bi_2Se_3$ by Volmer-Weber growth phenomena as indicated by the reciprocal lattice diffraction spotted pattern of Fig. S1. The spot pattern perfectly replicates the underlying twin populations of the $Bi_2Se_3$. And thus, two sets of dot patterns are obtained: one set for one twin domain labeled with red-colored indices, and another set corresponding to the other twin labeled with white-colored indices. In Fig. S1, two different RHEED patterns are presented. The pattern on the top right-side of Fig. S1 was obtained from $Bi_2Se_3$ grown on InP(111) with a high twin density, while the bottom right picture shows the RHEED pattern obtained for the $Bi_2Se_3$ grown with low-twin density. Mapping the relative intensity of the RHEED images enables relative quantification of the different twin populations.

By scanning along the lines indicated in Fig. S1 that connect diffraction spots corresponding to two different twin orientations, we can estimate the twin population ratio by the peak-height ratio at the (200) diffraction spots. In the nominal high twinned material we find a twinning ratio of 1:0.8, whereas the low twinned material we find a ratio of 1:0.2 and indicates a 4x reduction in twins.

Figures S2 and S3 show HR-XRD pole-scan measurements of the same high and low twinned material, respectively, and indicate a reduction of twin density consistent with the evidence from RHEED. In the pole scan measurement, the samples are oriented in the x-ray diffractometer to capture symmetric reflections from both the InP substrate ((002) type reflections) and the $Bi_2Se_3$ layer ((015) type reflections) grown on that. To capture the pole scan, the entire sample was rotated azimuthally about its circumference at the Bragg angles for the (015) and (002) reflections. The diffraction peaks from the InP (tall) and $Bi_2Se_3$ (short) were collected from both the high-twinned and low-twinned samples. The peaks from InP occur every 120º because it is a pure, untwinned single crystal substrate. Since the $Bi_2Se_3$ has twin defects, two sets of 120º periodic peaks corresponding to the twin domains are obtained with a 60º offset. In this way we see high-twinned $Bi_2Se_3$ has an effective six-fold symmetry. Due to the epitaxial alignment there is a convolution of the principal $Bi_2Se_3$ peaks with those of InP, while the twinned $Bi_2Se_3$ domains are in isolation at 60º intervals between the InP peaks.

Another quantitative measure of the twin density can be obtained by computing the peak-height ratio of the convoluted $Bi_2Se_3$ peak to the $Bi_2Se_3$ peak in isolation. Those two peaks correspond to different twin polarities. For the twinned (red) data set, the peak height ratio of convoluted to isolated peaks is roughly 1:0.84 and indicates that there is a roughly equal population of twins in both polarities. However, for the blue/right-side data set, we find a strikingly different result: the peak height ratio of convoluted to isolated peaks is roughly 1:0.22 and indicates that there approximately 4 times less twinning.

Quantification of the reduction in twin defects as measured by RHEED and by HR-XRD are in agreement, and they both consistently show a major reduction in twin defect density.

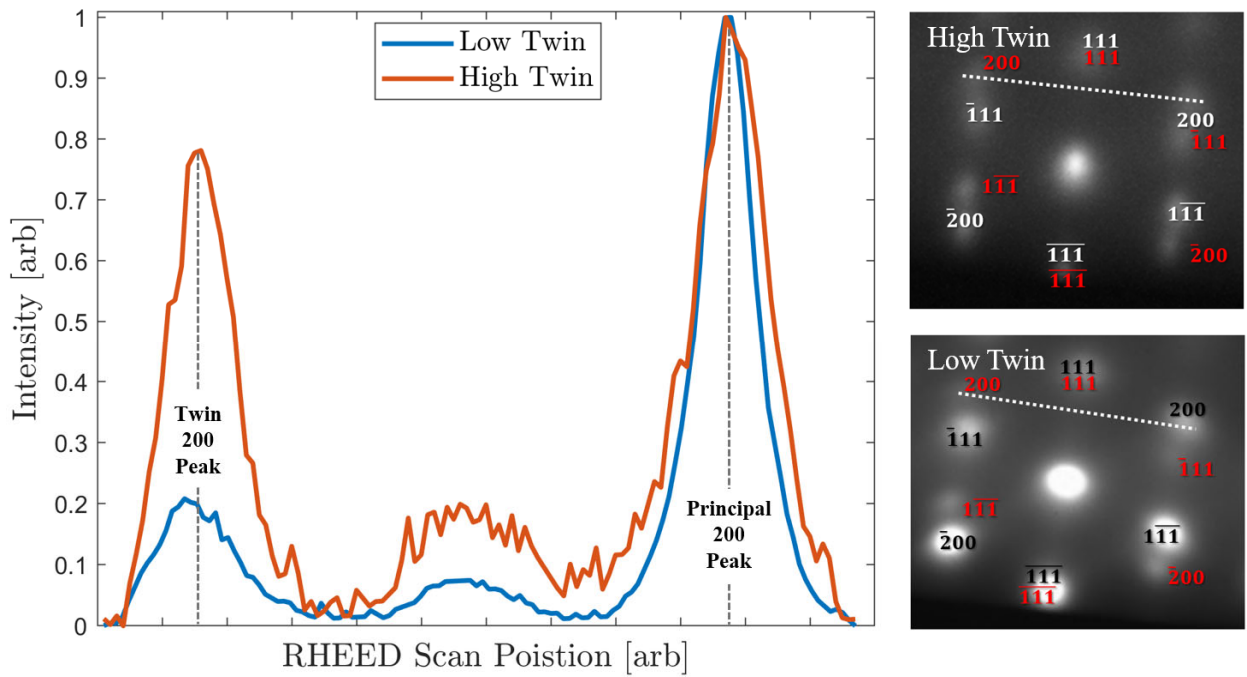

**Figure S1 | RHEED Scan.** RHEED intensity scan (left) of high (right-top) and low (right-bottom) twin $Bi_2Se_3$ samples. The relative intensities of the twinned 200 peaks are selected to measure twinning ratios. The low twinned material exhibits an approximate 1:0.2 ratio of principal to twin intensity, while the high shows 1:0.8.

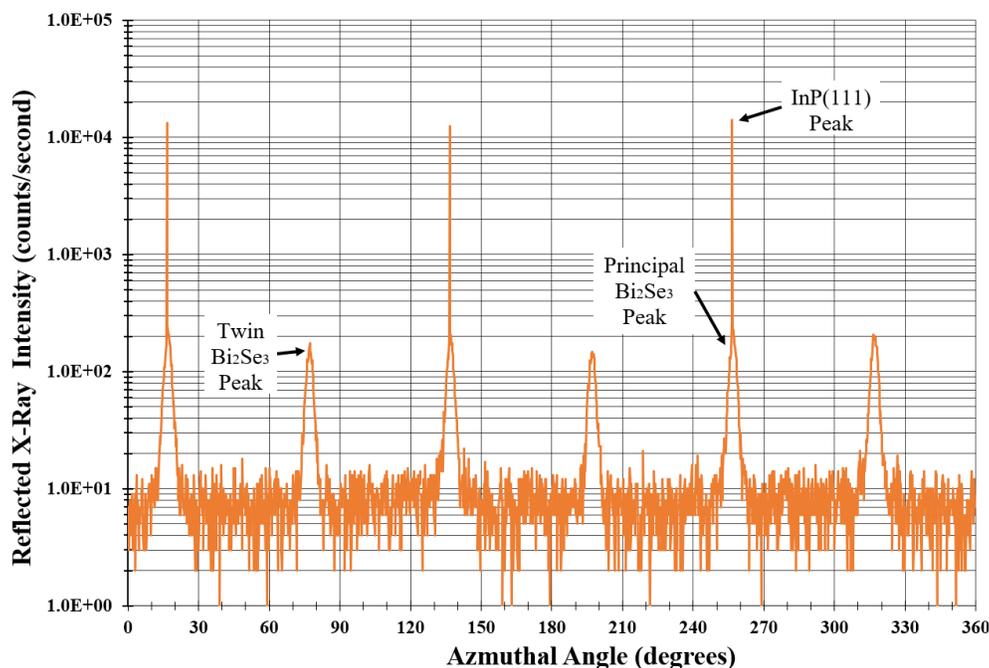

**Figure S2 | HR-XRD Pole-scan of Highly Twinned Bi$_2$Se$_3$.** The samples are oriented in the x-ray diffractometer to capture symmetric reflections from both the InP substrate ((002) type reflections) and the Bi$_2$Se$_3$ layer ((015) type reflections) grown on that. An approximate relative principal to twin intensity peak ratio of 1:0.84 is observed.

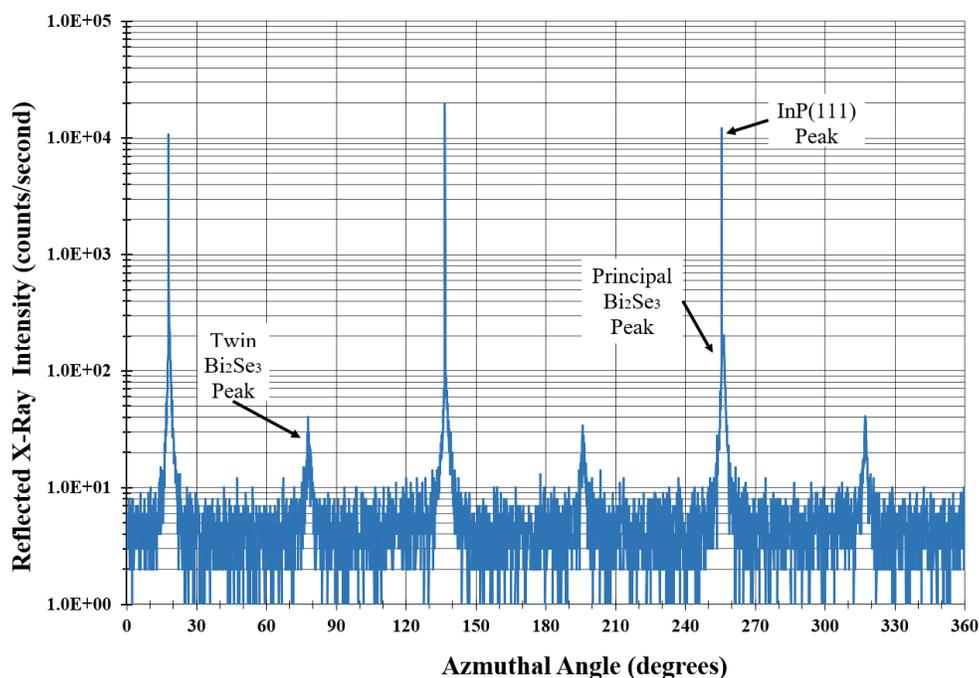

**Figure S3 | HR-XRD Pole-scan of Low Twinned Bi$_2$Se$_3$.** The samples are oriented in the x-ray diffractometer to capture symmetric reflections from both the InP substrate ((002) type reflections) and the Bi$_2$Se$_3$ layer ((015) type reflections) grown on that. An approximate relative principal to twin intensity peak ratio of 1:0.22 is observed.

## 2. Comparison of THz data on low and high twinned Bi$_2$Se$_3$ samples

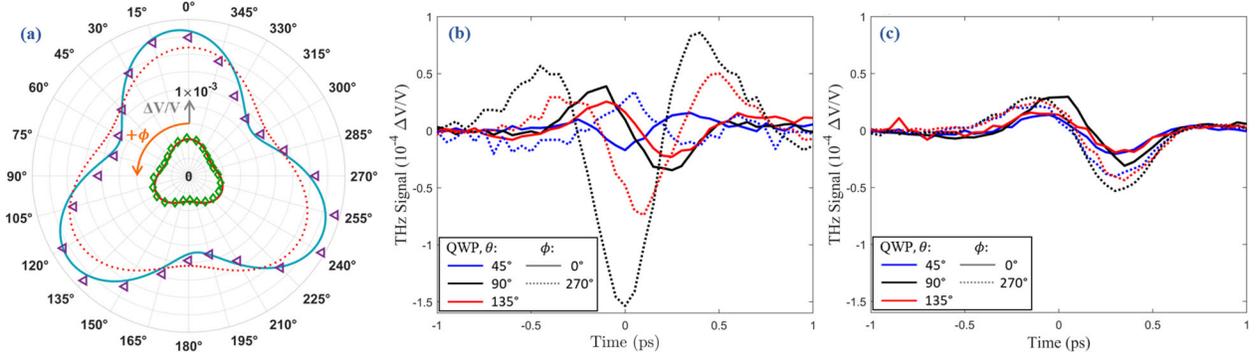

**Figure S4 | Comparison of THz emission from low and high twinned samples. (a)** Peak-to-peak amplitude of $S_{yz}$ as a function of $\phi$ for a low twinned sample (purple triangles) and a high twinned sample (green diamonds); the fitted function $\propto \cos(3\phi)$ (blue line and red lines, respectively) demonstrates the degree of three-fold crystallographic symmetry. For illustration, the high twinned fitted function is multiplied by 3.5 (dotted red line) and plotted for a direct comparison with the low twinned data. Of note is the smaller depth of modulation for the high twinned curve when compared with modulation of the low twinned sample, thus demonstrating the impact of twinning on crystal symmetries. **(b and c)** THz waveforms of $S_x$ at $\phi = 0°$ (solid lines) and $\phi = 270°$ (dashed lines) for QWP angles $\theta = 45°$ (right-hand circular; blue), $\theta = 90°$ (linear horizontal; black) and $\theta = 135°$ (left-hand circular; red), for **(b)** the low twinned sample and **(c)** the high twinned sample. Note, the inversion of sign between excitations of right and left-circular polarization for the low twinned sample for $\phi = 0°$; this polarization dependence is muted for $\phi = 270°$ as the response is dominated by the shift current when detection is aligned with the Se—Bi bond. For the high twinned sample in (c), we see minimal variation of $S_x$ as either the input polarization or the azimuthal orientation is varied due to suppression of photocurrents with an increase in twinning defect density.

## 3. Symmetry analysis of NLO tensors

The second order nonlinear optical (NLO) photocurrent response $j_i(\boldsymbol{q}, \omega)$ induced by an electric field pulse $E_i(\omega)$ centered at frequency $\omega$ and wavenumber $\boldsymbol{q}$ is given by:

$$j_i(\boldsymbol{q}, \omega) = \int d\omega' \left( \sigma_{ijk}(\omega' + \omega, -\omega') + q_l T_{iljk}(\omega' + \omega, -\omega') \right) \cdot E_j(\omega + \omega') E_k^*(\omega'), \quad (S1)$$

where the tensors $\sigma_{ijk}$ and $T_{iljk}$ encode the Photogalvanic Effect (PGE) and Photon Drag Effect (PDE) components of the NLO response; contributions proportional to higher orders in $q$ are ignored. Both $\sigma_{ijk}$ and $T_{iljk}$ must respect the underlying point group symmetries of Bi2Se3. In the DC limit ($\omega \to 0$), as discussed in other works, additional symmetries are imposed to guarantee reality of the photocurrent, but given the transient nature of the THz waveform, we do not impose this further constraint on our tensors. For a single twin domain, the surface of Bi2Se3 obeys the $C_{3v}$ point group symmetry, generated by three-fold rotations and single mirror plane, while the bulk obeys $D_{3d}$ which further includes inversion symmetry. A particular 'twin flavor' of $C_{3v}$ and $D_{3d}$ can be generated by

$$C_{3v} = \langle C_3, \sigma_{\hat{x}} \rangle, \quad D_{3d} \cong C_{3v} \otimes \langle i \rangle, \tag{S2}$$

where $C_3$ is a $\frac{2\pi}{3}$ rotation about the z-axis, $\sigma_{\hat{x}}$ is the mirror operator along the x-axis, and $i$ is the inversion operator. We can the compute the symmetry reduced forms[1,4,15,33] of $\sigma$ and $T$ by requiring for any generator $g$ of the point group: $[(g \otimes g \otimes g) \cdot \sigma]_{ijk} \equiv g_{ii'} g_{jj'} g_{kk'} \sigma_{i'j'k'} = \sigma_{ijk}$ and $[(g \otimes g \otimes g \otimes g) \cdot T]_{ijkl} = T_{ijkl}$. As inversion symmetry sends any odd-ranked tensor to its negative, i.e. $[(i \otimes i \ldots \otimes i) \cdot M]_{j_1 \ldots j_{2n+1}} \to -M_{j_1 \ldots j_{2n+1}}$, no bulk effects can contribute to PGE through $\sigma_{ijk}$, but as the PDE response is a fourth rank tensor it can include bulk effects. Instead of presenting the symmetry enforced forms of $\sigma$ and $T$ we simply show the final form of $j_i$ originating from PDE and PGE for our given setup details. We define our quarter wave plate (QWP) angle as $\theta$, and sample azimuthal angle $\phi$. Then, following Junck,[45] for an angle of incidence of 45°, the photon momentum of our incident electric field is given by $\boldsymbol{q} = \frac{q}{\sqrt{2}}(0,1,1)^T$, and after passing through the QWP the electric field vector is given by

$$\mathcal{E} \sim E_0 \left( -\frac{i \sin 2\theta}{\sqrt{2}}, \; -\frac{i}{2}(i + \cos 2\theta), \; \frac{i}{2}(i + \cos 2\theta) \right)^T. \tag{S3}$$

As we can see, the effect of a QWP on polarization is inherently two-fold periodic in $\theta$, and as such, when taking THz waveforms as a function of $\theta$ it is sufficient to take data only from $\theta = 0°$ to 180° and then artificially repeat it for display purposes. From here on we suppress expressions of frequency and any integration to simplify our equations. We compute the photocurrent responses as a function of $\phi$ by subjecting the symmetry reduced $\sigma$ and $T$ tensors to arbitrary z-axis rotations via: $\sigma_{ijk}(\phi) \equiv [(R_z(\phi) \otimes R_z(\phi) \otimes R_z(\phi)) \cdot \sigma]_{ijk}$ and

$T_{ijkl}(\phi) \equiv [(R_z(\phi) \otimes R_z(\phi) \otimes R_z(\phi) \otimes R_z(\phi)) \cdot T]_{ijkl}$. Then the PGE photocurrent response $\boldsymbol{j_G}$ is found to be:

$$j_{G,x} = \frac{E_0^2}{8}\Big(2\sqrt{2}\cos 3\phi \sin 4\theta\, \sigma_{xxy} - \sin 3\phi\, \sigma_{xxy}(1 - 3\cos 4\theta) + i\, 2\sqrt{2}\sin 2\theta\, (\sigma_{xxz} - \sigma_{xzx})$$
$$- \sqrt{2}\sin 4\theta\, (\sigma_{xxz} + \sigma_{xzx})\Big)$$

$$j_{G,y} = \frac{E_0^2}{8}\Big(\cos 3\phi\, \sigma_{xxy}(1 - 3\cos 4\theta) - 2\sqrt{2}\sin 3\phi \sin 4\theta\, \sigma_{xxy} - (\sigma_{xxz} + \sigma_{xzx})(3 + \cos 4\theta)\Big)$$

$$j_{G,z} = \frac{E_0^2}{8}(5\sigma_{zxx} + 3\,\sigma_{zzz} + \cos 4\theta\, (\sigma_{zzz} - \sigma_{zxx}))$$

Simplifying the expression we have the form presented in the main text:

$$\boldsymbol{j_G} = \begin{pmatrix} i\,\eta_2 \sin 2\theta + \eta_{x4}\sin 4\theta - \sin 3\phi\,(\tilde{\delta} + \tilde{\kappa}_4 \cos 4\theta) + \tilde{\eta}_4 \cos 3\phi \sin 4\theta \\ \delta_y + \kappa_{y4}\cos 4\theta + \cos 3\phi\,(\tilde{\delta} + \tilde{\kappa}_4 \cos 4\theta) - \tilde{\eta}_4 \sin 3\phi \sin 4\theta \\ \delta_z + \kappa_{z4}\cos 4\theta \end{pmatrix}. \quad (S4)$$

Note in the DC limit realness of $\boldsymbol{j}$ requires $\eta_{x2} \in \mathbb{R}$ and $\eta_{x4} = 0$. We see from the Fourier decomposed signals for different azimuthal angles of incidence in Fig. 5, $\eta_2$ changes sign as a function of angle. Additionally, we see in Fig. 3 the Fourier decomposition of fixed polarization the coefficients for $\cos 3\phi$ are always opposite. As $\cos 4\theta = -1$ & $\sin 4\theta = 0$ for circular polarized light, they cannot generate the sign difference in three-fold symmetry coefficients. Simple second order response thus predicts an azimuthally independent helical photocurrent response, with an azimuthally three-fold symmetric linear photocurrent response. It is clear by the above discussion this is not what we see in the experimental data, and so one needs to inspect the PDE response $\boldsymbol{j_D}$. To compactify expressions, we first present the different $\theta$ dependencies of $\boldsymbol{j_D}$:

$$\boldsymbol{j}|_{\sin 2\theta} = i\frac{qE_0^2}{4}\sin 2\theta \begin{pmatrix} T_{xxxx} - T_{xxyy} - 2T_{xyxy} + T_{xzxz} - T_{xzzx} + \cos 3\phi\,[T_{xxyz} - T_{xxzy}] \\ -\sin 3\phi\,[T_{xxyz} - T_{xxzy}] \\ 0 \end{pmatrix},$$

$$\boldsymbol{j}|_{\sin 4\theta} = \frac{qE_0^2}{8}\sin 4\theta \begin{pmatrix} T_{xxxx} - T_{xxyy} - T_{xzxz} - T_{xzzx} - \cos 3\phi\,[T_{xxyz} + T_{xxzy} - 2T_{xzxy}] \\ \sin 3\phi\,[T_{xxyz} + T_{xxzy} - 2T_{xzxy}] \\ -2\sin 3\phi\,T_{zxxy} \end{pmatrix},$$

$\boldsymbol{j}|_{\cos 4\theta}$

$$= \frac{qE_0^2}{8\sqrt{2}}\cos 4\theta \begin{pmatrix} \sin 3\phi\,(T_{xxyz} + T_{xxzy} - 3T_{xzxy}) \\ T_{xxxx} - 2T_{xxyy} - T_{xzxz} - T_{xzzx} + T_{xxzz} + \cos 3\phi\,[T_{xxyz} + T_{xxzy} - 3T_{xzxy}] \\ T_{zzzz} - T_{zxxz} - T_{zxzx} - T_{zzxx} - \cos 3\phi\,T_{zxxy} \end{pmatrix},$$

$\boldsymbol{j}|_{const}$

$$= \frac{qE_0^2}{8\sqrt{2}}\begin{pmatrix} \sin 3\phi\,(3T_{xxyz} + 3T_{xxzy} - T_{xzxy}) \\ 3T_{xxxx} + 2T_{xxyy} + 3T_{xxzz} - 3T_{xzxz} - 3T_{zxxz} + \cos 3\phi\,(3T_{xxyz} + 3T_{xxzy} - T_{xzxy}) \\ 5T_{zzxx} + 3T_{zzzz} - 3T_{zxzx} - 3T_{zxxz} - \cos 3\phi\,T_{zxxy} \end{pmatrix}.$$

These terms can be combined into the simplified expression in the main text:

$$\frac{\boldsymbol{j_D}}{q} \sim \begin{pmatrix} i\,\xi_2\sin 2\theta + \xi_{x4}\sin 4\theta + \sin 3\phi\,(\tilde{\Delta} + \tilde{\lambda}_4\cos 4\theta) + \cos 3\phi\,(i\,\tilde{\xi}_2\sin 2\theta + \tilde{\xi}_4\sin 4\theta) \\ \Delta_y + \lambda_{y4}\cos 4\theta + \cos 3\phi\,(\tilde{\Delta} + \tilde{\lambda}_4\cos 4\theta) - \sin 3\phi\,(i\,\tilde{\xi}_2\sin 2\theta + \tilde{\xi}_4\sin 4\theta) \\ \Delta_z + \lambda_{z4}\cos 4\theta + \cos 3\phi\,(\tilde{\Delta}_z + \tilde{\lambda}_{z4}\cos 4\theta) + \tilde{\xi}_{z4}\sin 3\phi\,\sin 4\theta \end{pmatrix} \quad (S5)$$

**Effects of Twinning and/or Six-fold Symmetry:**

Twinning mathematically implies we have an additional crystal mirrored along the $y$-axis present in our thin film. Our system then acquires on average an additional $c_{\hat{y}}$ symmetry, in turn making it six-fold symmetric. This is equivalent to saying the $C_{6v}$ point group describes our crystal surface, while $D_{6h}$ describes the bulk. We can either repeat the analysis of the tensors above by studying the forms symmetric under the generators of the new point groups, or equivalently enforce $c_{\hat{y}}$ symmetry on the three-fold rotationally symmetric tensors to reduce them further, as $R_z(\pi) = c_{\hat{y}} c_{\hat{x}}$, and $\sigma$ and $T$ are already invariant under $c_{\hat{x}}$. The equivalent approaches yield the following modified current densities:

$$\boldsymbol{j}_G = \begin{pmatrix} i\,\eta_2 \sin 2\theta + \eta_{x4} \sin 4\theta \\ \delta_y + \kappa_{y4} \cos 4\theta \\ \delta_z + \kappa_{z4} \cos 4\theta \end{pmatrix}, \quad (S6)$$

$$\frac{\boldsymbol{j}_D}{q} \sim \begin{pmatrix} i\,\xi_2 \sin 2\theta + \xi_{x4} \sin 4\theta \\ \Delta_y + \lambda_{y4} \cos 4\theta \\ \Delta_z + \lambda_{z4} \cos 4\theta \end{pmatrix}. \quad (S7)$$

We note yet another approach to obtaining the above results would be to compute the three-fold $\boldsymbol{j}_G$ and $\boldsymbol{j}_D$ responses for the other 'twin flavor' by applying $\sigma_{\hat{y}}$ to the $\sigma$ and $T$ operators computed for the original symmetry groups to obtain the twin tensors $\sigma^{twin}$ and $T^{twin}$. One can then compute $\boldsymbol{j}_G + \boldsymbol{j}_G^{twin}$ and $\boldsymbol{j}_D + \boldsymbol{j}_D^{twin}$ to arrive at the above expressions. If one is interested in intermediately twinned samples, this approach enables one to incorporate appropriate weights to the different twinning populations.

## 4. Azimuthal dependence of surface parallel photocurrents

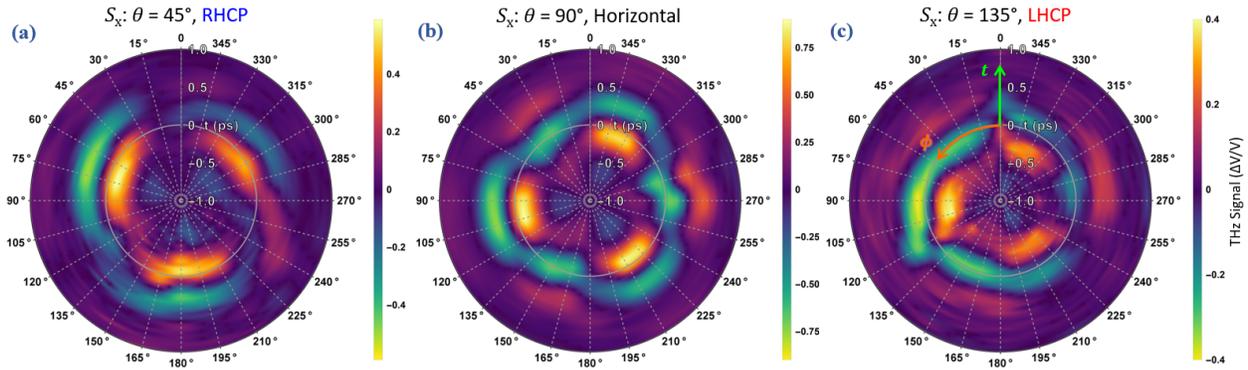

**Figure S5 | Azimuthal dependence of surface parallel photocurrents.** Vertical/surface THz signal, $S_x$, as a function of azimuthal angle, $\phi$, under excitation with: **(a)** Right-handed circular polarization (RHCP, $\theta = 45°$); **(b)** Linear horizontal polarization (Horizontal, $\theta = 90°$); **(c)** Left-handed circular polarization (LHCP, $\theta = 135°$). As indicated in (c), the temporal signals are plotted with time, $t$, along the radius and increasing $\phi$ proceeding counter clockwise. Note the clear three-fold symmetry evident in each plot.

## 5. Extended Fourier Analysis

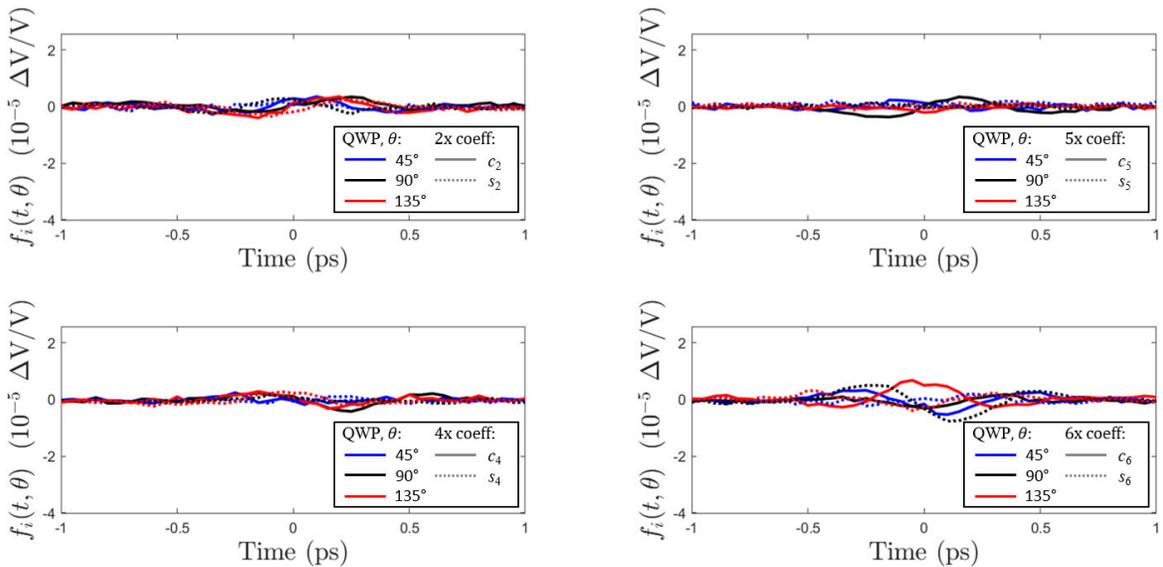

**Figure S6 | Additional Fourier decomposed THz contributions for sample azimuthal angle for $Bi_2Se_3$.** Two, four, five and six-fold modes are shown to compare to those in in Fig. 3 in the main text. Their contribution is significantly suppressed compared to the constant, one and three-fold modes predicted by symmetry analysis and substrate effects.

## 6. THz Waveforms for each QWP angle for from Fig. 4

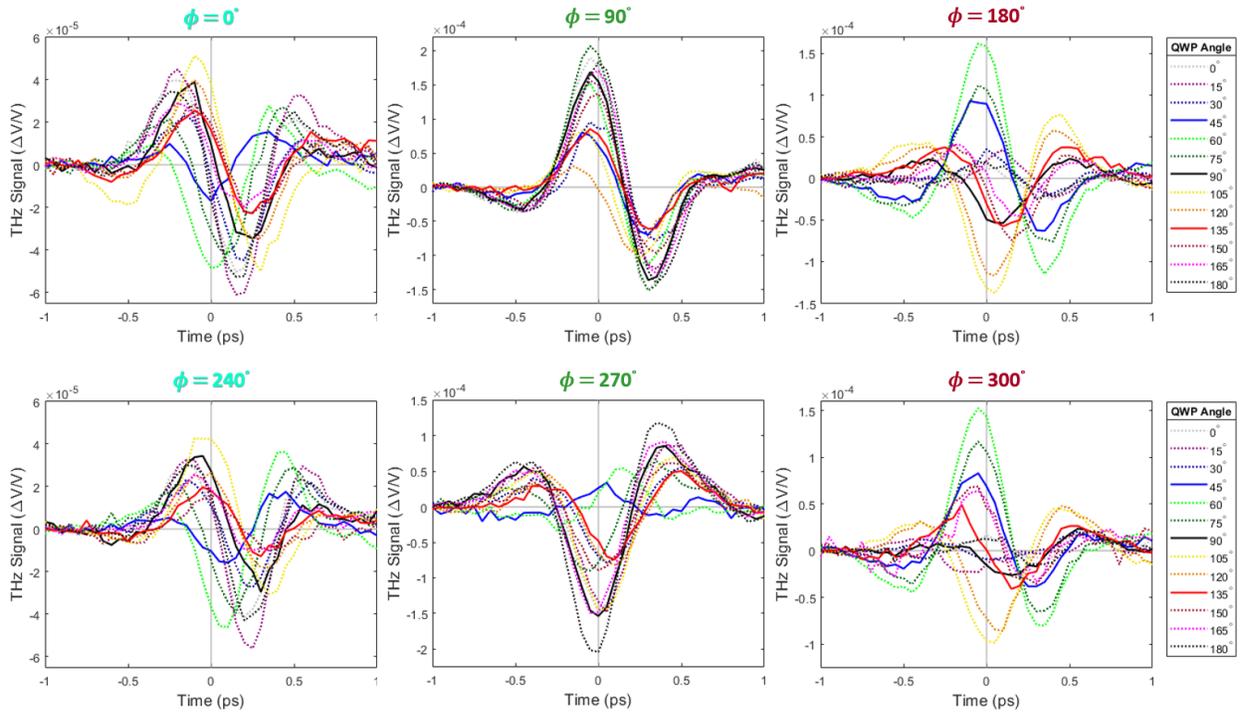

**Figure S7 | Full THz Waveforms for Helicity dependence of surface parallel photocurrents.** Full time-dependent $S_x$ waveforms (for radial plots in Fig. 4) as a function of the QWP angle, $\theta$, shown in legend, at discrete azimuthal angles indicated in each title.

## 7. InP Data

The InP(111) substrate also possesses three-fold rotational symmetry, and obeys similar symmetry analysis based calculation of polarization dependent photocurrents to $Bi_2Se_3$. Moreover, as InP breaks inversion symmetry, bulk bands could contribute to the PGE response. However, without strongly spin orbit split bands or topological Dirac surface states, microscopic mechanisms for generating PGE are strongly suppressed in InP(111) when compared to $Bi_2Se_3$. To ensure our observations do not result from the substrate, we carried out TDTS measurements on a bare InP(111) substrate for a selection of the same azimuthal angles as shown in Fig. 5. Radial plots as a function of $\theta$ are shown in Fig. S8. We also performed the same Fourier decomposition to extract periodicities with respect to $\theta$, which is presented in Fig. S8. The photocurrents generated in InP(111) are an order-of-magnitude smaller than $Bi_2Se_3$. Moreover, the thin films of $Bi_2Se_3$ we study will significantly absorb the light before it reaches the InP(111), further reducing any photocurrent generated in the buried substrate. As such, we can safely conclude the observed three-fold symmetric photocurrents are principally a result of the low twinned $Bi_2Se_3$.

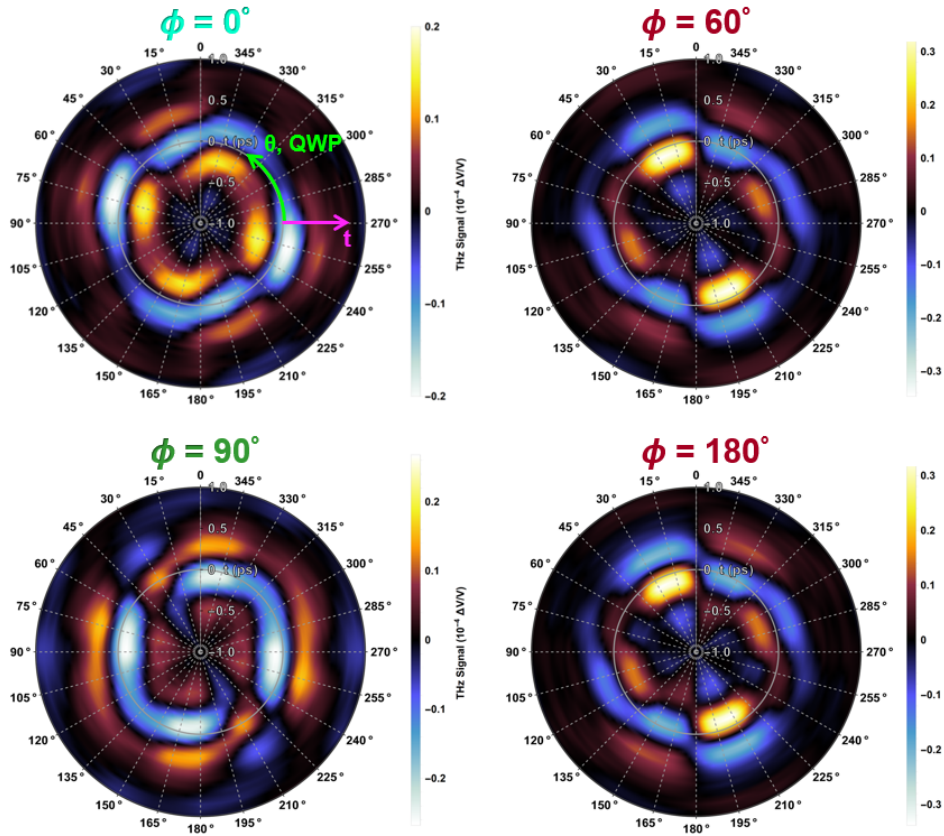

**Figure S8 | Helicity dependence of InP(111) substrate surface parallel photocurrents.** Radial plots of $S_x$ waveforms measured from TDTS on a bare InP(111) substrate as a function of the QWP angle, $\theta$, with time on the radial axis and $\theta$ on the polar axis; each radial plot is displayed on an independent color scale (as shown). The raw data is taken for QWP angles $\theta = 0°$ to $180°$ in $15°$ steps, and then piecewise interpolated with a third-degree polynomial to achieve a finer plot resolution. The interpolated data set is then repeated for $\theta \in (180°, 360°]$ to achieve a full radial plot (QWP polarization states are inherently two-fold periodic). Strong three-fold symmetry is seen by the nearly identical $\phi = 60°$ and $180°$ data. We see strong two-fold QWP periodicity in $\phi = 60°, 180°$ and $270°$, while four-fold periodicity dominates for $\phi = 0°$. This is phenomenologically similar to the results on low twinned $Bi_2Se_3$, but with an order of magnitude smaller currents, different relative four-fold/two-fold strengths and translated polarization of the maximum response.

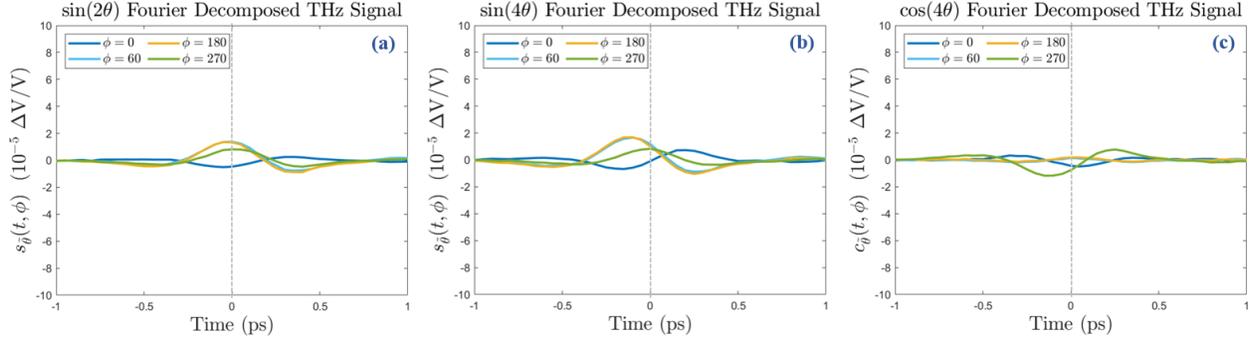

**Figure S9 | Fourier decomposed THz contributions for QWP angle for InP.** The set of time dependent Fourier coefficients corresponding to **(a)** $\sin(2\theta)$, **(b)** $\sin(4\theta)$, and **(c)** $\cos(4\theta)$ behavior of THz waveforms measured for photocurrent excitations on a bare InP(111) offcut substrate are shown for several sample angles. The azimuthal angles $\phi$ are defined so that $\phi = 0°$ is the same orientation of the InP(111) substrate in both these measurements and those on $Bi_2Se_3$ in the main text sample azimuthal angles. The signals obey three-fold symmetry, i.e. $\phi = 60°$ and $\phi = 180°$ are nearly identical as expected from the three-fold rotational symmetry of InP(111). We observe that the InP(111) waveforms are roughly an order of magnitude smaller than those generated by $Bi_2Se_3$, and when further accounting for absorption of light by the epitaxial layer of $Bi_2Se_3$ in the main experiment, InP(111) cannot be responsible for the magnitude of response observed.

Page 34 of 34